\documentclass[journal,onecolumn,12pt, draftclsnofoot]{IEEEtranTIE}

\usepackage{xcolor,soul,framed} 

\colorlet{shadecolor}{yellow}

\usepackage[pdftex]{graphicx}
\graphicspath{{../pdf/}{../jpeg/}}
\DeclareGraphicsExtensions{.pdf,.jpeg,.png}

\usepackage[cmex10]{amsmath}

\usepackage{array}
\usepackage{mdwmath}
\usepackage{mdwtab}
\usepackage{eqparbox}
\usepackage{url}
\usepackage{booktabs}
\usepackage{stfloats}
\usepackage{subfigure}
\usepackage{cite}
\usepackage{multirow}

\begin{document}
\bstctlcite{IEEEexample:BSTcontrol}
    \title{Automatic Receiver Tracking and Power Channeling  for Multi-Transmitter Wireless Power Transfer}
  \author{
 Xiaojie~Dang,
       Prasad~Jayathurathnage,
       S.~A.~Tretyakov,
       and~C.~R.~Simovski
}


\maketitle

\begin{abstract}

Free positioning of receivers is one of the key requirements for many wireless power transfer (WPT) applications, required from the end-user point of view. However,  realization of stable and effective wireless power transfer for  freely positioned receivers is technically challenging task because of the requirement of complex control and tuning. In this paper, we propose a concept of automatic receiver tracking and power channeling for multi-transmitter WPT systems using uncoupled transmitter and uncoupled repeaters. Each transmitter-repeater pair forms an independent power transfer channel providing an effective link for the power flow from the transmitter to the receiver. The proposed WPT system is capable of maintaining stable output power with constant high efficiency regardless of the receiver position and without having any active control or tuning. The proposed concept is numerically and experimentally verified by using a four-transmitter WPT system in form of a linear array. The experimental results show that the efficiency of the proposed WPT system can reach  94.5\% with a variation less than 2\% against the receiver position.

\end{abstract}

\begin{IEEEkeywords}
Wireless power transfer, overlapping coils, multiple transmitter coils, free positioning.
\end{IEEEkeywords}
\IEEEpeerreviewmaketitle

\section{Introduction}

\IEEEPARstart{W}{ireless} power transfer (WPT) technology has attracted increased attention in many applications such as electric vehicles, medical implant devices, consumer electronics, and industrial applications, due to its numerous advantages. 
 However, there are several technical challenges to overcome for wide and successful deployment of the WPT technology, which make the WPT technology an important  research topic. One of these critical problems is that  the transferred power and the energy efficiency is rapidly reduced as the WPT distance increases.   Therefore, the limitation of the WPT range is a challenge in most of the WPT applications.

One of the effective methods to increase the transmission range is the use of intermediate repeaters, which relay the magnetic energy from transmitter (Tx) to receiver (Rx), enhancing the power transfer capability \cite{Ahn2013, Hua2019}. From the user view point, one of the key challenges in WPT applications is  the requirement of perfect alignment between the Tx and the Rx. For example, in case of consumer electronics applications,  such as wireless charging of mobile phones, laptops, tablet PCs, etc., free positioning of  devices is  most desirable, so that one can put the mobile device in an arbitrary position still ensuring  wireless charging. With emerging new WPT applications such as WPT enabled kitchen appliances, WPT enabled furniture, or WPT for lighting applications, the requirement of free positioning is becoming more and more demanding. 

A number of approaches have been proposed to enable free positioning of  WPT receivers over a large area. All these proposals can be broadly categorized into two types: the use of large transmitting coil structures to generate a uniform magnetic field distribution over a large area, or the use of multiple transmitters to cover a large area. The first approach is simple and straightforward, however, such approaches suffer from inferior efficiency due to  very weak coupling between the Tx and Rx. For example, Liu et.~al. employs a  combination of concentrated winding and spiral winding to design the transmitter and realize a uniform magnetic field \cite{Liu2008}. However, the efficiency is about only 50\% in the presence of one receiver. 

Therefore, the use of multiple transmitter coils is the preferred method for most of the applications \cite{Jow2013, Zhong2011, Mi2016, AbdollahMirbozorgi2014, Mirbozorgi2016, Jolani2015}. However, the excitation of  Tx coils has become a complex problem. The simplest way is to excite all the Txs simultaneously \cite{AbdollahMirbozorgi2014, Mirbozorgi2016, Jolani2015,Kim2012}. However, such simultaneous and homogeneous activation is not optimal because the power contribution from each Tx to the Rx depends on the corresponding mutual coupling. Transmitter coils with strong coupling to the receiver transfer more power. In fact, the  optimal ratio of Tx coil currents depends on the mutual inductances between each Tx-Rx pair  \cite{jaya_optCurrentDistribution,Huh2018}. Therefore, simultaneous activation of many transmitters incur unnecessary power losses in the transmitters that are weakly coupled with the receiver. 

There have been several proposals to deactivate or suppress the currents in weakly coupled transmitters such as the use of additional on-off switches \cite{Hui2013,Li2019}, switchable compensation circuits \cite{pacini2017load}, and additional switchable coupled-loops \cite{Kim2019}. Therefore, the requirement of the Rx position detection, the complex control approaches, and additional tuning circuits have become the critical concerns for selective activation of multi-Tx WPT. The complexity and the cost of the system will further increase if the receiver is continuously moving (for example, in electric vehicle charging) due to the requirement of additional control circuits to track the movement.

This paper proposes a multi-Tx WPT system capable of automatic receiver tracking and optimal power channeling through nearby coils without the use of any sensing or control circuits. The suggested approach got inspiration from the known idea of so called ``superlens'' based on two parallel arrays of small resonant objects \cite{Maslovski2004}. In that device, the resonators in each array are decoupled from each other, realizing point-to-point channels for energy for arbitrarily positioned sources. This functionality is critically important for WPT systems for freely positioned receivers, which we introduce here. The proposed WPT system consists of multiple Tx-repeater pairs, which are referred to as \emph{independent channels} in this paper. The proposed WPT system realizes a uniform high-efficiency and stable output power throughout all Rx positions within the coverage area. The physical arrangement of the coils can be in any form as long as the following requirements are met: 1.~All the coils except ones in the same channel are uncoupled from each other, and 2.~For a given channel, coupling between the Tx and the Rx is small as compared to the coupling between the repeater and the Rx. Even though the ideal conditions may not be perfectly satisfied in practical implementations, we show that the concept of power channeling can be achieved with a proper design and optimization. A prototype implementation is numerically and experimentally verified in this paper. A transmitter array and a repeater array are formed by using a set of partially overlapping spiral coils with very weak cross coupling between them. The experimental results show that stable high efficiency can be maintained for arbitrary the receiver positions without using any active tuning or control.

The paper is organized as follows. In Section~II, we analyze  single Tx-Rx WPT systems including the conventional 2-coil WPT system and the 3-coil WPT system with a repeater, and find a simplified expressions for the currents in each coils, providing the basis for understanding the proposed automatic receiver tracking and power channeling. In Section~III, we discuss and analyze the concept of the proposed WPT scheme with respect to the coils currents, output power stability, and the efficiency characteristics. Section~IV presents experimental results to verify the property of  automatic  switching between  the corresponding power channels with small variations of the output power and efficiency. Finally, this paper concludes with a summary of the results in Section~V.

\section{Wireless Power Transfer System with  a Single Transmitter and Receiver}

\subsection{The 2-coil WPT system}
We start our discussion with the analysis of  the  2-coil WPT system that includes a single Tx-Rx pair. The equivalent circuit of the WPT system is shown in Fig.~\ref{circuit_diagram}, where two coils are connected to series compensation capacitors $C_{\rm{tx}}$ and $C_{\rm{rx}}$. Tx is excited by a voltage source $V_{\rm{s}}$ and the Rx is connected to an electrical load with an impedance of $R_{\rm{L}}$. Both resonators have the same resonance frequency $\omega_0=1/\sqrt{L_{\rm{tx}} C_{\rm{tx}}}=1/\sqrt{L_{\rm{rx}}C_{\rm{rx}}}$ where $L_{\rm{tx}}$ and $L_{\rm{rx}}$ are inductances of Tx ad Rx, respectively. By employing the fundamental-harmonic linear approximation, we can write the system equation as
\begin{equation}
    \begin{bmatrix}
         Z_{\rm{tx}} & {\rm{j}}\omega M_{\rm{tx\text{-}rx}}
        \\ {\rm{j}}\omega M_{\rm{tx\text{-}rx}}  & R_{\rm{L}}+Z_{\rm{rx}} 
    \end{bmatrix}
         \begin{bmatrix}
         I_{\rm{tx}} \\ I_{\rm{rx}} 
    \end{bmatrix}=
         \begin{bmatrix}
         V_{\rm{s}}\\ 0  
     \end{bmatrix}
\end{equation}
where $Z_{i}=R_{i}+{\rm{j}}X_{i} $, $X_{i}=\omega L_{i}-1/C_{i} $ ($i=\rm{tx, rx}$), $R_{\rm{tx}}$ and $R_{\rm{rx}}$ are the parasitic resistances of the coils, and $ M_{\rm{tx\text{-}rx}}$ is the mutual inductance between the coils. One can easily find the currents in Tx and Rx ($I_{\rm{tx}}$ and $I_{\rm{rx}}$, respectively) at the resonant frequency $\omega_0$ as
\begin{equation}\label{i_no_rp}
\begin{split}
   &I_{\rm{tx}}=\frac{V_{\rm{s}}\left(R_{\rm{rx}}+R_{\rm{L}}\right)}{\omega_0^2 M_{\rm{tx\text{-}rx}}^{2}+R_{\rm{tx}}\left(R_{\rm{rx}}+R_{\rm{L}}\right)},\\
   &I_{\rm{rx}}=\frac{-{\rm j} \omega_0 M_{\rm{tx\text{-}rx}}V_{\rm{s}}}{\omega_0^2 M_{\rm{tx\text{-}rx}}^{2}+R_{\rm{tx}}\left(R_{\rm{rx}}+R_{\rm{L}}\right)}.
\end{split}
\end{equation}
With the intention of understanding the multi-Tx WPT system, let us analyze the characteristics of $I_{\rm{tx}}$ and $I_{\rm{rx}}$ with respect to the change of $M_{\rm{tx\text{-}rx}}$. For a given source voltage $V_{\rm s}$, the maximum current in the Rx flows  when $M_{\rm{tx\text{-}rx}}=\sqrt{R_{\rm{tx}}\left(R_{\rm{rx}}+R_{\rm{L}}\right)}/\omega_0$, which essentially corresponds to a particular Rx position. When the Rx  moves away from this position, the Rx current decreases regardless of the Tx current. However, the Tx current increases with the decrease of $M_{\rm{tx\text{-}rx}}$. Decreasing $M_{\rm{tx\text{-}rx}}$ implies that Rx moves away from the Tx. When $M_{\rm{tx\text{-}rx}}$ is small, Tx current is much higher than the Rx current, which leads to very low efficiency and a possibility of short circuit of the source. 

In a multi-Tx WPT scenario, transmitters that are farther from the receiver should have smaller currents compared to those closer to the receiver. However, the characteristics of $I_{\rm{tx}}$ and $I_{\rm{rx}}$ of 2-coil WPT are completely opposite to what we want to achieve. Therefore, one should continuously detect the position of the Rx and control the Tx currents, which require additional sensing circuits and complex control approaches. Hence, the simple 2-coil WPT system is not promising for extending to multi-Tx WPT systems.

\subsection{WPT System with a repeater}
\begin{figure}[!t]
\centering 
\subfigure[]{
\includegraphics[width=0.5\columnwidth]{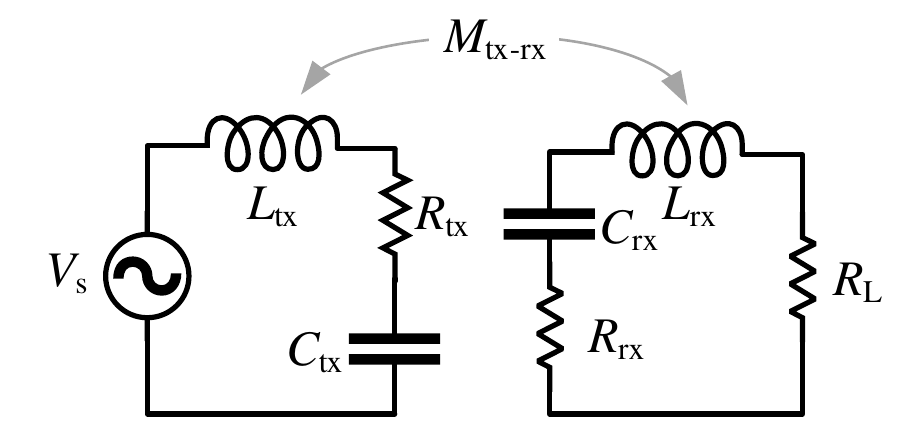}\label{circuit_diagram} }  
\subfigure[]{
\includegraphics[width=0.7\columnwidth]{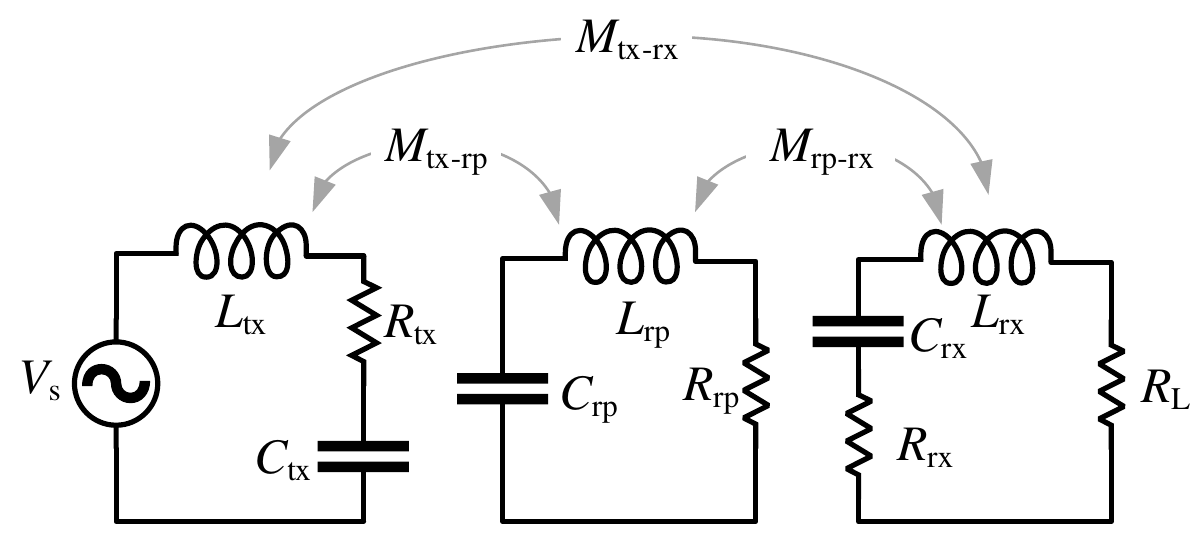}\label{circuit_diagram2} }
\caption{Equivalent circuit model of the wireless power transfer systems; (a)~Two-coil system, and (b)~WPT system with a repeater.}
\end{figure}

Now, let us analyze a 3-coil WPT system with an additional intermediate repeater (Rp). The equivalent circuit of the 3-coil WPT system is shown in Fig.~\ref{circuit_diagram2}. The resonance frequency of the repeater is identical to the other two: $\omega_0=1/\sqrt{L_{\rm{tx}} C_{\rm{tx}}}=1/\sqrt{L_{\rm{rp}}C_{\rm{rp}}}=1/\sqrt{L_{\rm{rx}}C_{\rm{rx}}}$ where the subscript rp refers to the repeater. 


We employ Kirchhoff's voltage law to define the system equation as
\begin{equation}\label{kvl}
\begin{bmatrix}
 Z_{\rm{tx}} & {\rm{j}}\omega M_{\rm{tx\text{-}rp}} & {\rm{j}}\omega M_{\rm{tx\text{-}rx}} 
\\ {\rm{j}}\omega M_{\rm{tx\text{-}rp}} & Z_{\rm{rp}} & {\rm{j}}\omega M_{\rm{rp\text{-}rx}} 
\\ {\rm{j}}\omega M_{\rm{tx\text{-}rx}} & {\rm{j}}\omega M_{\rm{rp\text{-}rx}}  & R_{\rm{L}}+Z_{\rm{rx}} 
\end{bmatrix}
 \begin{bmatrix}
 I_{\rm{tx}}\\ I_{\rm{rp}} \\ I_{\rm{rx}} 
\end{bmatrix}=
 \begin{bmatrix}
 V_{\rm{s}}\\ 0 \\ 0 
 \end{bmatrix}
\end{equation}
 After solving (\ref{kvl}), one can find the current in each coil. Assuming identical Tx and repeater coils (i.e $R_{\rm{tx}}=R_{\rm{rp}}=R$) and that the parasitic resistance of the Rx is much smaller compared to the load impedance ($R_{\rm{L}} \gg R_{\rm{rx}} \rightarrow R_{\rm{L}} + R_{\rm{rx}} \approx R_{\rm{L}}$), the coil currents at the resonant frequency can be written as
\begin{equation}\label{3coilcurrents_full}
\begin{split}
   &I_{\rm{tx}}=\dfrac{(M_{\rm{rp\text{-}rx}}^2\omega_0^2+RR_{\rm L})V_{\rm{s}}}{A}\\
   &I_{\rm{rp}}=\dfrac{-{\rm{j}}V_{\rm{s}}\omega_0 M_{\rm{tx\text{-}rp}}R_{\rm L}\left(1-{\rm{j}}\frac{\Delta}{2}\right)}{A}\\
   &I_{\rm{rx}}=\dfrac{-{\rm{j}} V_{\rm{s}}(M_{\rm{tx\text{-}rx}}R\omega_0)-{\rm{j}}M_{\rm{rp\text{-}rx}}M_{\rm{tx\text{-}rp}}\omega_0^2}{A}
\end{split}
\end{equation}
where
\begin{equation}\label{Delta}
\begin{split}
& A=R^2R_{\rm L}+R\omega_0^2( M_{\rm{tx\text{-}rx}}^{2}+M_{\rm{rp\text{-}rx}}^{2})+M_{\rm{tx\text{-}rp}}^{2}R_{\rm{L}}\omega_0^2\left(1-{\rm{j}}\Delta\right),\\
& \quad \quad \quad\quad\quad\quad \text{and}~\Delta  =\dfrac{2M_{\rm{tx\text{-}rx}}M_{\rm{rp\text{-}rx}}\omega_0}{M_{\rm{tx\text{-}rp}}R_{\rm{L}}}.
\end{split}
\end{equation}
When the mutual inductance $ M_{\rm{tx\text{-}rx}} \ll M_{\rm{tx\text{-}rp}} $, the term $\Delta\ll 1$. This mutual inductance relation can easily be realized when the distance between Tx and the repeater is smaller than the distance between Tx and Rx. Under the condition of $\Delta\ll 1$ and neglecting the parasitic resistance of the coils, i.e., $R \approx 0$, \eqref{3coilcurrents_full} can be simplified to
 \begin{equation}\label{3coilcurrents_smpl2}
 \begin{split}
   &I_{\rm{tx}}=\frac{M_{\rm{rp\text{-}rx}}^{2}V_{\rm{s}}}{M_{\rm{tx\text{-}rp}}^{2}R_{\rm{L}}}\\
   &I_{\rm{rp}}=\frac{-{\rm{j}} V_{\rm{s}}}{\omega_0 M_{\rm{tx\text{-}rp}}}\\
   &I_{\rm{rx}}=\frac{-M_{\rm{rp\text{-}rx}}V_{\rm{s}}}{M_{\rm{tx\text{-}rp}}R_{\rm{L}}}
\end{split}   
\end{equation}
We can observe an interesting characteristic in (\ref{3coilcurrents_smpl2}). The Rx current is proportional to $M_{\rm{rp\text{-}rx}}$, while the Tx current is proportional to $M_{\rm{rp\text{-}rx}}^2$, which is totally different from the characteristics of the 2-coil WPT system. When the mutual coupling between the repeater and the Rx increases, the currents in both Tx and Rx increase. This is in fact an essential and desired property for multi-Tx WPT. We see that when the receiver current is small (the receiver is far away or loaded with a high-resistance load), the current in the primary coil is suppressed while the repeater coil is still strongly excited. In the limiting case  $R_{\rm L}\rightarrow \infty$, the Tx current tends to zero.  This practically useful feature has an interesting analogy with superlenses based on the use of resonant particles. Also in that device, the current in the first resonator (here we excite it directly by a voltage source) is suppressed, while the current in the second resonator (our repeater) is enhanced, transferring the fields behind the arrays.  
Importantly, the repeater current is non-singular even in the limit of negligible losses and absent receiver.

Let us also note that the load voltage $V_{\rm L}$ can be approximated as $\left|V_{\rm L} \right| \approx \left|\omega_0 M_{\rm{tx\text{-}rx}} I_{\rm{tx}} +  \omega_0 M_{\rm{rp\text{-}rx}} I_{\rm{rp}}\right|$. The direct contribution to the load voltage from the Tx is much smaller than that of  the repeater. This is because of two reasons: firstly, the mutual inductance  $M_{\rm{tx\text{-}rx}}$ is much less than $M_{\rm{rp\text{-}rx}}$, secondly, the voltage contributions from the transmitter and the repeater are in 90 degrees out of phase.

\begin{figure}
  \begin{center}
  \includegraphics[width=3.5in]{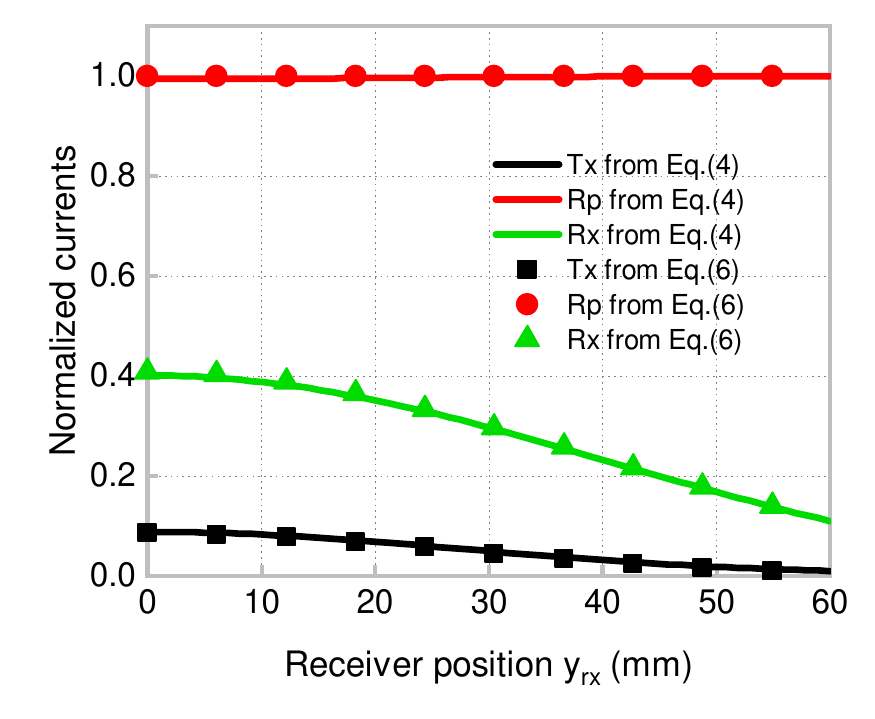}\\
  \caption{Normalized coil currents in the Tx, repeater (Rp), and the Rx of the 3-coil WPT systems versus the receiver position in $y$-direction.}\label{onechannel}
  \end{center}
\end{figure}

Before moving to the multi-Tx WPT scheme, let us consider a numerical example of 3-coil WPT system with three identical coils (the coil parameters are defined in Table~\ref{Table.coil_parameters}) in order to validate the above analytical results and approximations. The distances between the Tx to the repeater and the repeater to the Rx are $10$~mm and $50$~mm, respectively. In this example, $\Delta\approx 0.1$, which meets the requirement of (\ref{3coilcurrents_smpl2}). The normalized coil currents using the full analytical form in \eqref{3coilcurrents_full} are compared with the simplified derivation in \eqref{3coilcurrents_smpl2}. The results are shown in Fig.~\ref{onechannel}, which verifies the accuracy of the simplification. We can also observe that the Tx current decreases with Rx misalignment while keeping the repeater current almost constant. Therefore, we can exploit this property of the 3-coil WPT to achieve automatic receiver tracking and power channeling in a multi-Tx WPT, which will be discussed in detail in the following section.

\begin{table}[!t]
\caption{Parameters of Coils in WPT system}
\centering
\label{Table.coil_parameters}
\begin{tabular}{@{}ll@{}}
\toprule\toprule
Parameter                    &  Value \\ \midrule
Diameter                     & 100 mm \\
Number of turns              &  8      \\
Pitch between turns          & 4 mm   \\
Diameter of wire& 2 mm  \\

\bottomrule\bottomrule
\end{tabular}
\end{table}

\section{Multi-Transmitter and Multi-Repeater System}

\subsection{System Configuration}
Let us consider a multi-Tx WPT system consisting of multiple transmitter-repeater pairs. Each pair is referred to as a \emph{channel}. We also assume that coupling between all the coils except ones in the same channel are negligibly small. Now, the multi-Tx WPT system comprises a set of independent 3-coil WPT channels. This is again analogous to the main principle of superlenses based on resonant arrays. Also in those  devices, the resonators in each array should be as weakly coupled to each other as possible, so that any  distribution of currents over the array resonators corresponds to a resonant mode at the operational frequency. This property ensures creation of independent, point-to-point energy channels across the ``lens''. Note that, according to the above discussion, the current in a particular Tx decreases  when the Rx is moved away from it.
\begin{figure}[!t]
  \begin{center}
  \includegraphics[width=3.5in]{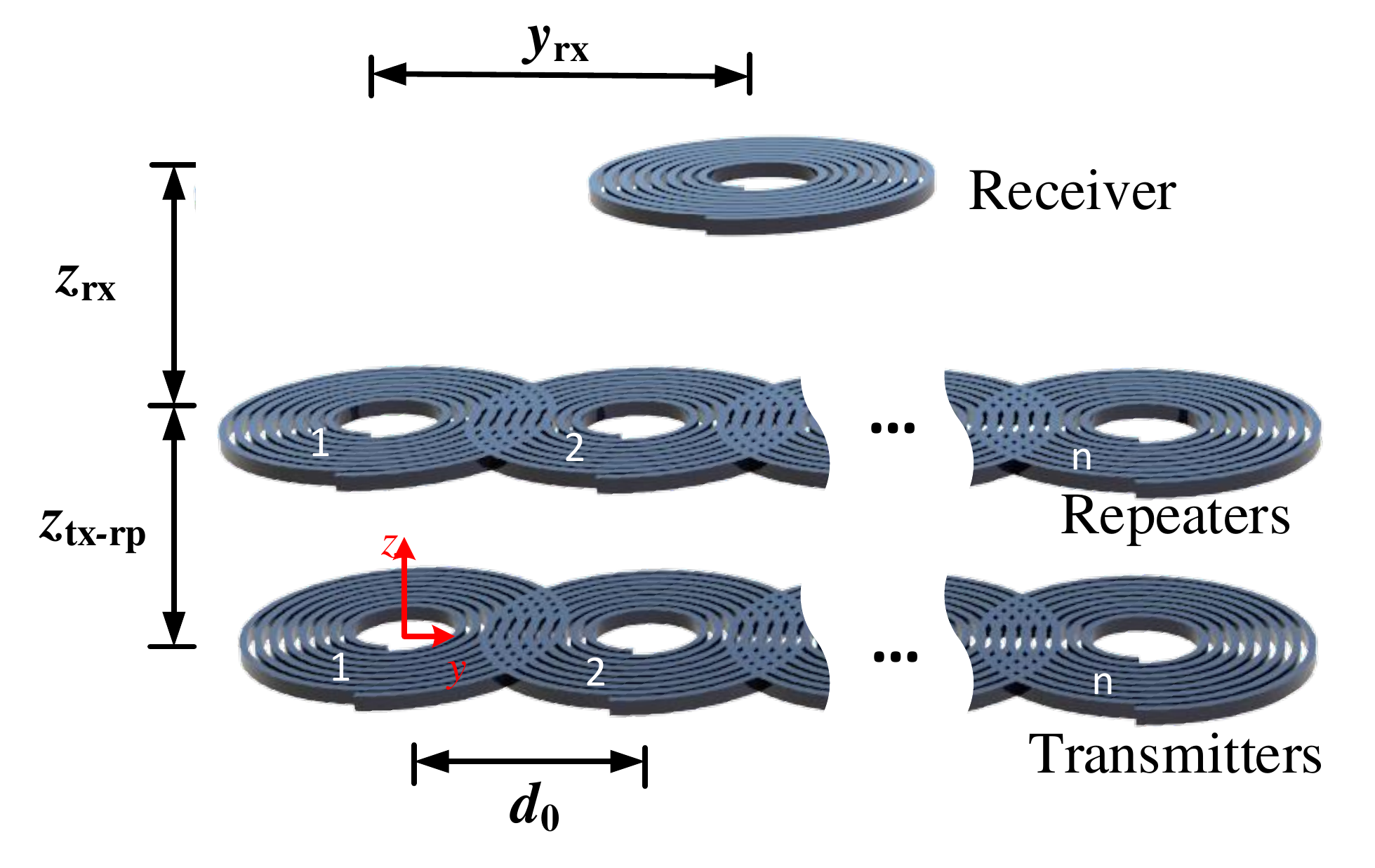}\\
  \caption{The coil arrangement of the multi-Tx WPT system.}\label{multitx}
  \end{center}
\end{figure}

The geometrical configuration can be in any form, for example, 1D linear array, 2D coil pad, or even a 3D arrangement. The only requirement is to have negligible cross coupling between the channels. For example, a coil arrangement of a 1D linear array is shown in Fig.~\ref{multitx} and the Rx can be positioned (or moved) along the array. We design the Tx and repeater arrays with partially overlapping coils to ensure zero mutual coupling between the adjacent coils. All the coils are chosen to be identical circular spirals with the design parameters given in Tables \ref{Table.coil_parameters} and \ref{Table.coil_arragement}. The working frequency of the WPT system is chosen to be at $1$~MHz and a series tuning capacitor is connected with each coil to tune the resonant frequencies to the working frequency. The transmitters connected in parallel and are excited by a voltage source.

\begin{table}[!t]
\caption{Parameters of Coil Arrangement}
\centering
\label{Table.coil_arragement}
\begin{tabular}{@{}ll@{}}
\toprule\toprule
Parameters      &  Values \\ \midrule
Distance between two adjacent coils ($d_0$ )     & 57.55 mm \\
Distance between transmitter and repeater array ($z_{\rm{tx\text{-}rp}}$ )  &   10 mm      \\
Distance between repeater array and receiver ($z_{\rm{rp\text{-}rx}}$ )   & 20 mm \\
                                                                     & 50 mm   \\
 \bottomrule\bottomrule
\end{tabular}
\end{table}

\subsection{Analysis of the Multi-Tx WPT System}
For the analysis of the multi-Tx WPT system, we assume that the WPT system consists of $n$  independent channels. Let us start with the analytical derivation of Rx  current $I_{\rm{rx}}$ as a function of currents in the Txs and the repeaters. Using the equivalent circuit analysis, the Rx current $I_{\rm{rx}}$ is expressed as
 \begin{equation}\label{nchannelIRx}
  I_{\rm{rx}}=-\frac{\sum\limits_{i=1}^n {\rm{j}}\omega_0 M_{{\rm{tx}}i\rm{\text{-}rx}}I_{{\rm{tx}}i}+\sum\limits_{i=1}^n {\rm{j}}\omega_0 M_{{\rm{rp}}i\rm{\text{-}rx}}I_{{\rm{rp}}i}}{R_{\rm{L}}+R}
\end{equation}
where the subscript $i$ refers to the $i^{\rm th}$ Tx or repeater coil. 
The output power at the load $P_{\rm out}$ is defined as 
\begin{equation}\label{pout_fullform}
P_{\rm out}=\left | I_{\rm{rx}} \right |^2 R_{\rm{L}},
\end{equation}
and the power transfer efficiency $\eta$ of the WPT system is defined as
\begin{equation}\label{pte}
    \eta=\frac{\left | I_{\rm{rx}} \right |^2 R_{\rm{L}}}{\sum\limits_{i=1}^{n}\left | I_{{\rm{tx}}i} \right |^2 R+\sum\limits_{i=1}^{n}\left | I_{{\rm{rp}}i} \right |^2 R+\left | I_{\rm{rx}} \right |^2 \left ( R+R_{\rm{L}} \right )}.
\end{equation}
 We assume that the channels are independent and the mutual coupling between Txs and the Rx is much weaker than that of the corresponding repeaters and the Rx. As we discussed in Section~II, the power contribution directly from Txs to the Rx is very small compared to that from repeaters to the Rx. Therefore, we can approximate the currents in each coil individually by extending (\ref{3coilcurrents_smpl2}). This way, we can obtain the output power of the system as
  \begin{equation}\label{pout}
  \begin{split}
P_{\rm out}&=\frac{V_{\rm s}^2}{k_{{\rm{tx}}{\rm{\text{-}rp}}}^2R_{\rm L}}\left(\sum\limits_{i=1}^{n}k_{{\rm{rp}}i{\rm{\text{-}rx}}}\right)^2,\\
k_{{\rm{tx\text{-}rp}}}&=\frac{M_{\rm tx\text{-}rp}}{\sqrt{L_{\rm{tx}} L_{\rm{rp}}}},\\
k_{{\rm{rp}}i\rm{\text{-}rx}}&=\frac{M_{{\rm{rp}}i\rm{\text{-}rx}}}{\sqrt{L_{{\rm rp} i} L_{\rm{rx}}}}
\end{split}
\end{equation}
where  $k_{{\rm{tx\text{-}rp}}}$ and $k_{{\rm{rp}}i\rm{\text{-}rx}}$ are the coupling coefficients between Tx-repeater pairs in each independent channel and between the $i^{\rm th}$ repeater and the Rx, respectively.

In order to characterize the efficiency, we separate the loss contributions from the different segments of the system, i.e, the Tx array, the repeater array, and the Rx. To this end,  we define power loss ratios for each segment ($\xi_{\rm{tx,rx,rp}}=P_{\rm{tx,rx,rp}}/P_{\rm out}$) as the ratios between the power losses in each segment and the output power ($P_{\rm out}$). Substituting (\ref{3coilcurrents_smpl2}) into (\ref{pte}), the efficiency of the system can be simplified to
\begin{equation}\label{ptenew2}
\begin{split}
\eta &=\frac{1}{1+\xi_{\rm{tx}}+\xi_{\rm{rx}}+\xi_{\rm{rp}}},\\
\xi_{\rm{tx}} &= \dfrac{\left(\sum\limits_{i=1}^{n}k_{{\rm{rp}}i{\rm{\text{-}rx}}}^4\right)}{ \Gamma k_{\rm{tx\text{-}rp}}^2\left(\sum\limits_{i=1}^{n}k_{{\rm{rp}}i{\rm{\text{-}rx}}}\right)^2},\\
\xi_{\rm{rp}} &=\dfrac{n \Gamma}{Q^2 \left(\sum\limits_{i=1}^{n}k_{{\rm{rp}}i{\rm{\text{-}rx}}}\right)^2},\\
\xi_{\rm{rx}} &=\dfrac{1}{\Gamma}, \quad \text{and \quad} \Gamma=\dfrac{R_{\rm L}}{R},
\end{split}
\end{equation}
where  $Q=\omega_0 L/R$ is the unloaded quality factor of the coil, $\Gamma$ is the ratio between the load resistance ($R_{\rm L}$) and the coil resistance ($R$), which is generally much greater than unity, i.e., $\Gamma \gg 1$ .

To maximize the efficiency in \eqref{ptenew2}, one should minimize all the loss ratios. The loss ratio of the Tx array $\xi_{\rm tx}$ is much smaller compared to the other two loss ratios because of the fourth-order coupling summation term in the numerator and $\Gamma \gg 1$. The highest loss contribution comes from the repeaters, which can be decreased by increasing the sum of the coupling terms, or by increasing the coil quality factor.
In order to evaluate the output power and the efficiency characteristics at different Rx positions, let us consider an Rx that moves along the Tx array, as illustrated in Fig.~\ref{multitx}. When the Rx moves, the mutual coupling between each repeater to the Rx $k_{{\rm{rp}}i{\rm{\text{-}rx}}}$ changes, but the sum terms, $\sum_{i=1}^{n}k_{{\rm{rp}}i{\rm{\text{-}rx}}}$ and $\sum_{i=1}^{n}k_{{\rm{rp}}i{\rm{\text{-}rx}}}^4$ remain almost constant because a reduction of one coupling coefficient is compensated by an increase of another. Therefore, a nearly constant output power and high efficiency can be achieved. The output power and efficiency characteristics are further evaluated using a numerical example of 4-Tx array in the following section.

\subsection{Numerical verification}

\begin{figure}[!t]
 \begin{center}
 \includegraphics[width=0.6\columnwidth]{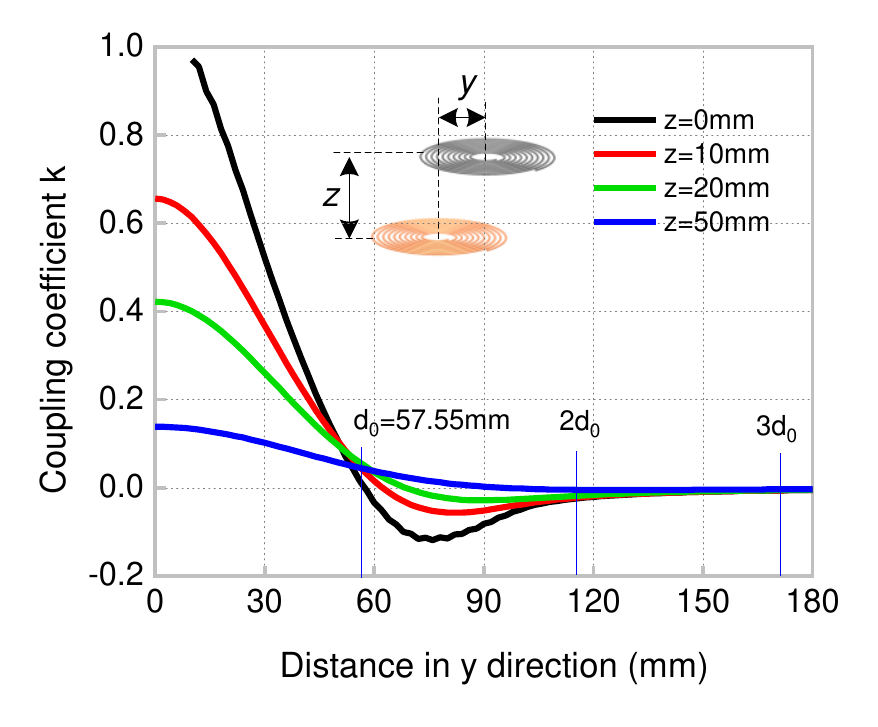}\\
 \caption{Coupling coefficient between two coils versus their displacement in $y$-direction.}\label{k_vs_y}
 \end{center}
\end{figure}

We consider a 4-Tx WPT system as an example to validate the properties of automatic receiver tracking and power channeling. The design parameters of the coils and the system are given in Table~\ref{Table.coil_parameters} and Table~\ref{Table.coil_arragement}, respectively. The coupling coefficients between the coils are numerically calculated using the circular loop approximation as proposed in \cite{MLcalculations}. The coupling coefficient variations with respect to the lateral displacement between the WPT coils are illustrated in Fig.~\ref{k_vs_y}. The adjacent coils are spaced $57.55$~mm away from each other, which is the uncoupling distance ($d_0$), corresponding to zero mutual inductance (see the line marked  $z=0$ in Fig.~\ref{k_vs_y}). It can also be observed from Fig.~\ref{k_vs_y} that the coupling coefficients for non-adjacent Tx and repeater coils (indicated as $2d_0$ and $3d_0$) are negligibly small (less than $0.026$) and the assumption of independent channels is valid. 


\begin{figure}[!t]
 \begin{center}
 \includegraphics[width=0.6\columnwidth]{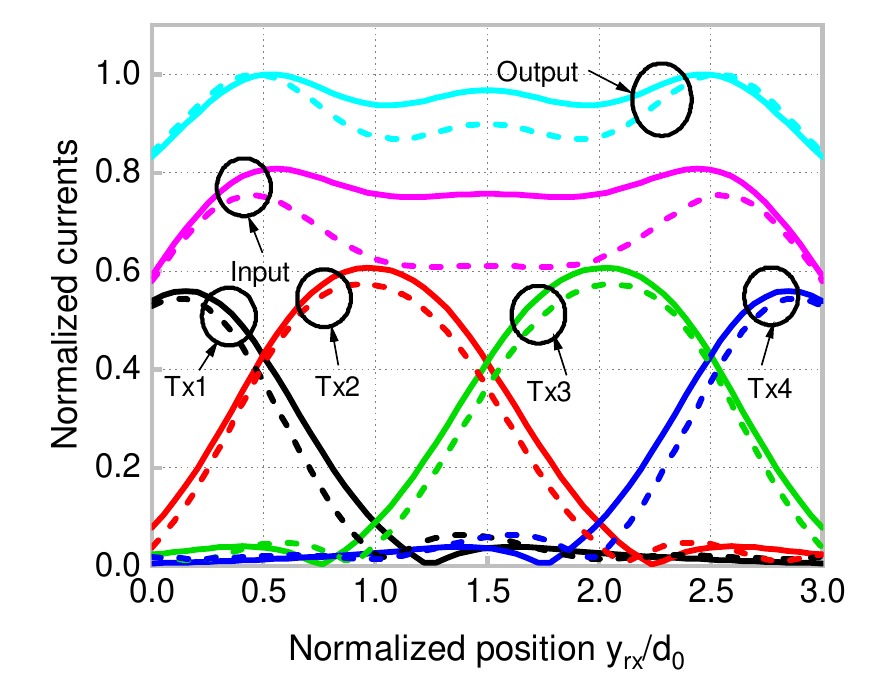}\\
 \caption{The variation of coil currents versus the normalized position to the receiver ($z_{\rm{rp\text{-}rx}}=20$~mm). Lines correspond to the practical WPT system including all the cross coupling terms. Dashed lines correspond to the ideal scenario with the assumption of zero-coupling between the channels.}\label{currents_4coil}
 \end{center}
\end{figure}
The numerically calculated currents in Txs and Rx with respect to the Rx position are shown in Fig.~\ref{currents_4coil}. In order to make the positions of channels clearer, we normalize the $y$-coordinates of the receiver position ($y_{\rm rx}$) by the separation  between two adjacent channels as $y_{\rm rx}/d_0$. The receiver position varies from the center of Channel~$1$ (i.e., aligned to Tx1) to the center of Channel $4$ (i.e., aligned to Tx4). The continuous lines in Fig.~\ref{currents_4coil} correspond to the practical setup when all the cross-coupling terms are considered, and the dashed lines correspond to the ideal setup when all the cross-coupling terms are assumed to be zero. For the ideal structure, there is no coupling between any two coils belonging to two different channels, while for the practical set-up, two adjacent coils of transmitters or repeaters are positioned with uncoupling  displacement $d_0$, but the mutual coupling still exists between non-adjacent coils. As shown in Fig.~\ref{currents_4coil}, the currents of the practical design (line) is closely following the results for the ideal condition (dashed line). The results verify that the cross-coupling terms are negligibly small and the solutions follow the ideal profile.

Let us now analyze the Tx current profiles at different Rx positions. For example, when the Rx moves away from Tx1, the current in Tx1 decreases while the current in Tx2 gradually increases. When the Rx is aligned to Tx2, the current in Tx2 reaches its peak, and the currents in all the other Txs are negligibly small. When the position of the Rx is in-between two Txs, the currents in the nearest Txs are identical meaning that the two Txs equally contribute to the power transfer. When the Rx moves along the Tx array, the currents in the four transmitter coils are adjusted automatically to transfer power optimally to the Rx. In the view of power channeling, four independent channels are automatically switched in conjunction with the position of the receiver. The input current, which is the sum of all the Tx currents remains almost constant. Most importantly, we can notice from Fig.~\ref{currents_4coil} that the variation in the Rx current (the output current) is very small. This result means that we achieve almost constant output power regardless of Rx position.  

\begin{figure}[!t]
 \begin{center}
 \includegraphics[width=0.6\columnwidth]{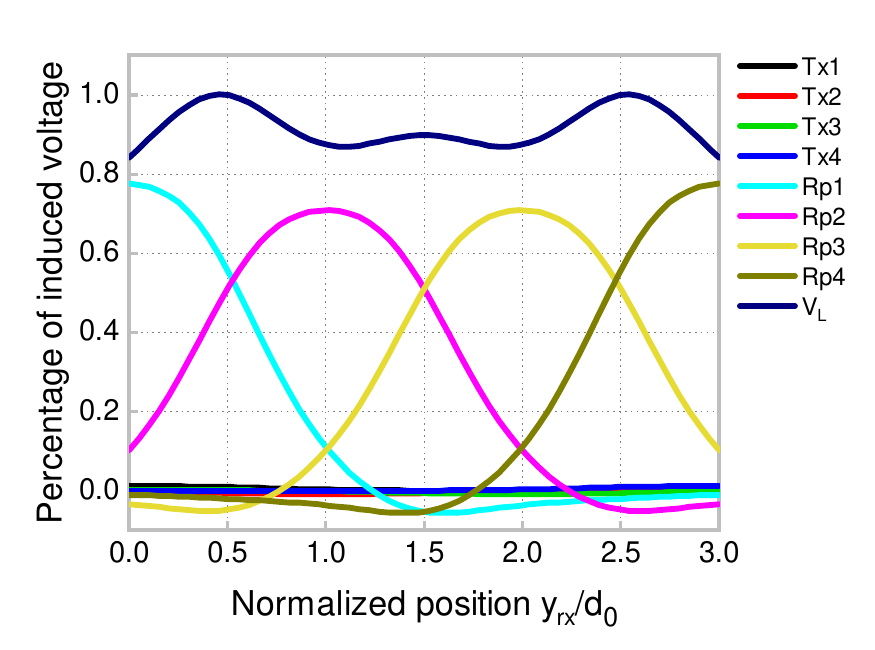}\\
 \caption{The percentage contribution to the load voltage from the Tx currents and the repeater currents ($z_{\rm{rp\text{-}rx}}=20$~mm).}\label{percentage}
 \end{center}
\end{figure}

In order to further analyze the characteristics of the output power, the percentage contribution to the induced voltage at the receiver (which is approximately equal to the load voltage $V_{\rm L}$) from the Txs and the repeaters with respect to the Rx position is illustrated in Fig.~\ref{percentage}. Similar to the 3-coil WPT system that we discussed in previous section, the main contribution to the induced voltage at the Rx coil is from the repeaters and the direct contribution from the Tx currents is very small. In fact, power flows from the Txs to the repeaters, and then it channels to the Rx through the repeater. The repeater currents are almost constant, which makes the Rx current remain almost constant throughout the Tx array. The position-invariant current in receiver can improve the system misalignment tolerance with almost constant output power.

\begin{figure}
 \begin{center}
 \includegraphics[width=0.6\columnwidth]{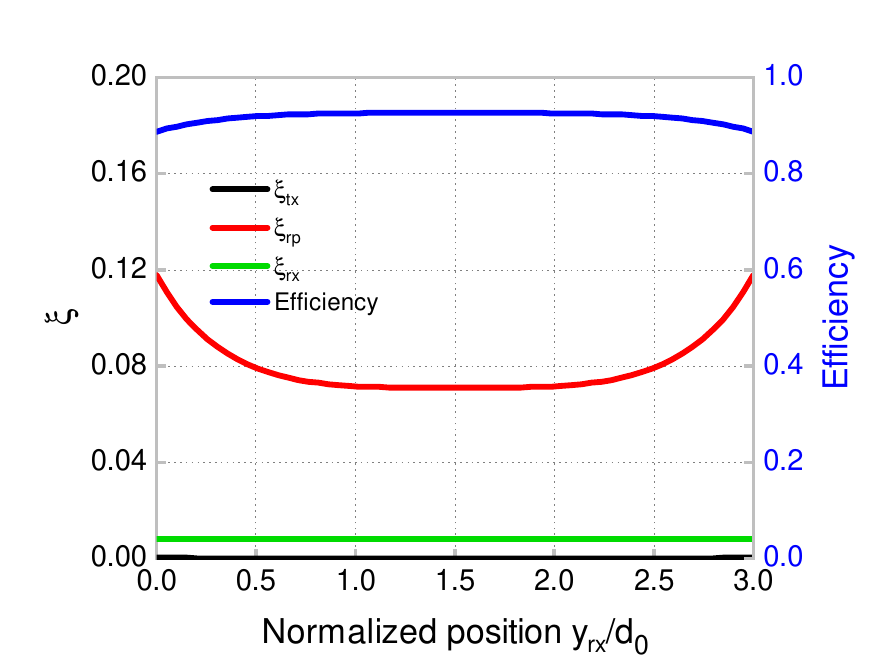}\\
 \caption{Power loss ratio and the efficiency variations against the receiver position in $y$-direction (coordinates are normalized to the channel separation $d_0$, and $z_{\rm{rp\text{-}rx}}=50$~mm).}\label{loss_pte}
 \end{center}
\end{figure}

Next, let us move to the analysis of the system efficiency with respect to the Rx position. The variation of loss ratios and the efficiency for different Rx positions are illustrated in Fig.~\ref{loss_pte}. The overall efficiency is maintained above $90\%$. As we discussed in the  previous section, the highest loss contribution is from the repeater array while the loss ratio of the Tx array is very small. Therefore, the results verify that the proposed power channeling approach not only maintains constant power at the receiver, but also achieves high efficiency. 


\section{Experimental Verification}
\begin{figure}[!t]
 \begin{center}
 \includegraphics[width=0.6\columnwidth]{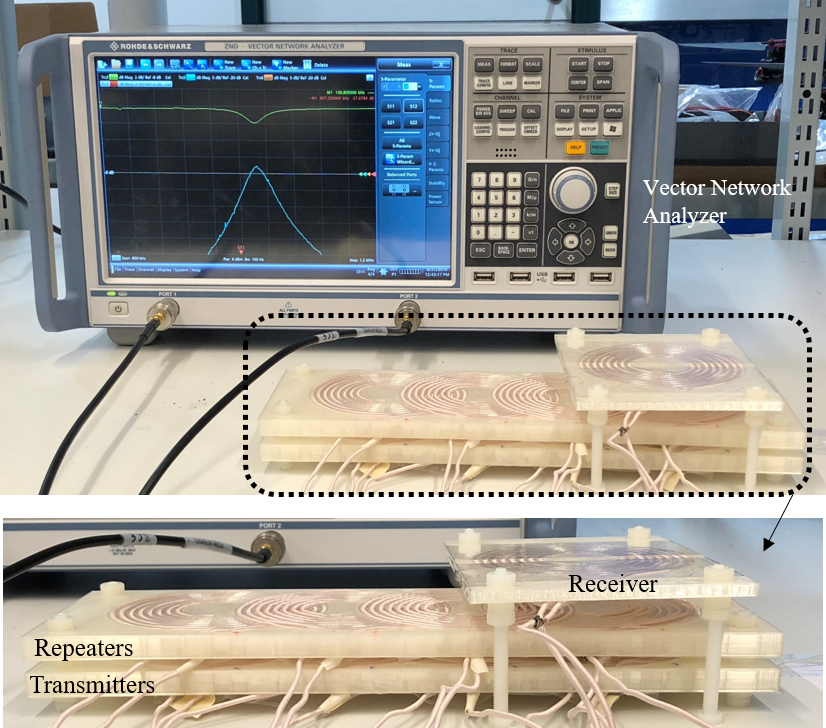}\\
 \caption{The experimental setup.}\label{photo}
 \end{center}
\end{figure}

Now, we move to the experimental verification of the proposed WPT system with the capability of automatic receiver tracking and power channeling. The experimental setup is shown in Fig.~\ref{photo}, and includes four transmitters, four repeaters, and one receiver. All the coil parameters are chosen to match Table~\ref{Table.coil_parameters}, and the coils are wounded using the Litz wire. The self-inductance and resistance of the coils were measured using an Agilent4284A LCR meter, and the measured results are shown in Table~\ref{Table.coil_measure}.  Series capacitors are added to each coil to make the coil resonant close to $1$~MHz, and the experimentally measured resonant frequencies are given in Table~\ref{Table.coil_measure}. In order to make the mutual inductance of each pair of adjacent coil to be zero, the distance between two neighbour coil is fixed to $57.55$~mm, based on the results in Fig.~\ref{k_vs_y}. The measured mutual inductance of two adjacent coils was less than $75$~nH, which is negligible compared to the main coupling terms. The parameters of the experimental setup are matched to the WPT system described in the previous section and presented in Table~\ref{Table.coil_arragement}. The receiver position moves along the $y$-axis from the center of Tx1 (i.e., Channel 1) to the center of Tx4 (Channel 4).


\begin{table}[!t]
\caption{Measured Parameters of the WPT Coils}
\centering
\label{Table.coil_measure}
\begin{tabular}{@{}lccc@{}}
\toprule\toprule
  & \multicolumn{1}{l}{Inductance ($\mu$H)} & \multicolumn{1}{l}{Resistance (m$\Omega$)} & \multicolumn{1}{l}{$f_0$ (MHz)} \\ \midrule
Tx1 & 4.64   & 55     & 0.9972             \\
Tx2 & 4.69   & 53     & 0.9968             \\
Tx3 & 4.69  & 55     & 0.9990            \\
Tx4 & 4.90  & 67     & 0.9990            \\
Rp1 & 4.70  & 49     & 1.0024            \\
Rp2 & 4.58  & 52     & 0.9930            \\
Rp3 & 4.61  & 57     & 0.9988            \\
Rp4 & 4.73  & 65     & 1.0024            \\
Rx & 4.47  & 42     & 1.0020            \\ \bottomrule\bottomrule
\end{tabular}
\end{table}

The measurements are carried out using a vector network analyzer (VNA) where all the Txs are connected in parallel to Port~$1$ and the Rx is connected to Port~$2$. The measured two-port network parameters are post-processed in Keysight Advanced Design System tool \cite{ADSsoftware} using a similar approach as presented in \cite{kim2014ADS}. 

\begin{figure}[!t]
 \begin{center}
 \includegraphics[width=0.8\columnwidth]{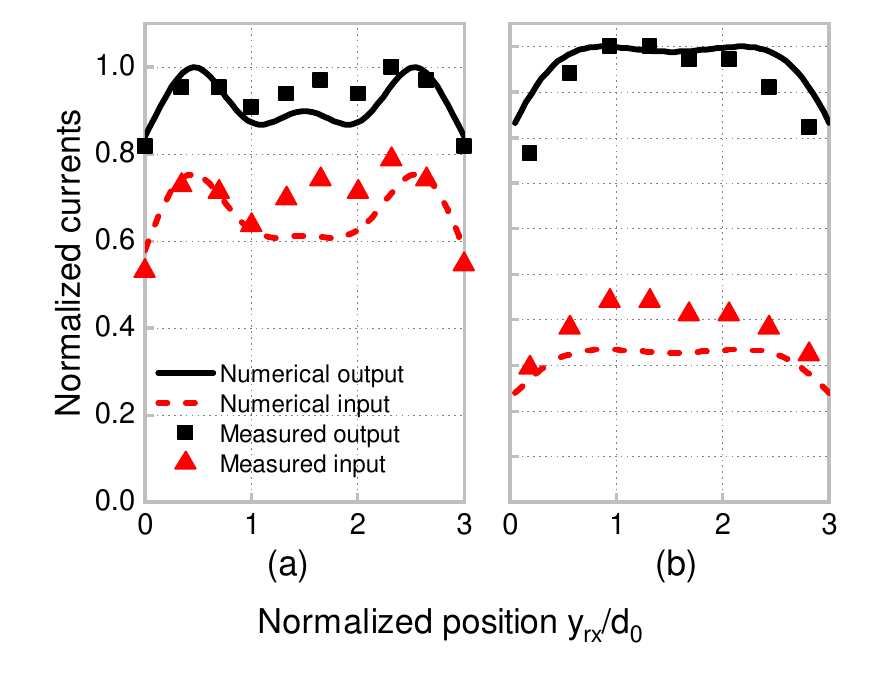}\\
 \caption{Normalized input current (i.e., the sum of all Tx currents) and the output current against the position of the receiver when the $z$-distance between the repeater array and the receiver is (a)~20 mm, and (b)~50 mm.}\label{currentgraph}
 \end{center}
\end{figure}

Fig.~\ref{currentgraph} shows the normalized input current (i.e., the sum of all the Tx currents) and the normalized receiver current (the output current) with respect to the receiver position $y_{\rm rx}/d_0$. Due to the constraint of two ports during the experiment, the individual currents in each transmitter coil cannot be obtained directly. However, it is noted that the measured input current is in very good agreement with the numerical results, which validates the theoretical analysis. When the Rx is closer to a certain Tx, the current in this Tx increases, which can be considered as an activated channel for power transfer. When the position of the receiver changes, four transfer channels are switched on and off accordingly. Therefore, the output current is almost constant when the receiver moves along the Tx array, which verifies the theoretical basis of the power channeling.

\begin{figure}
 \begin{center}
 \includegraphics[width=0.8\columnwidth]{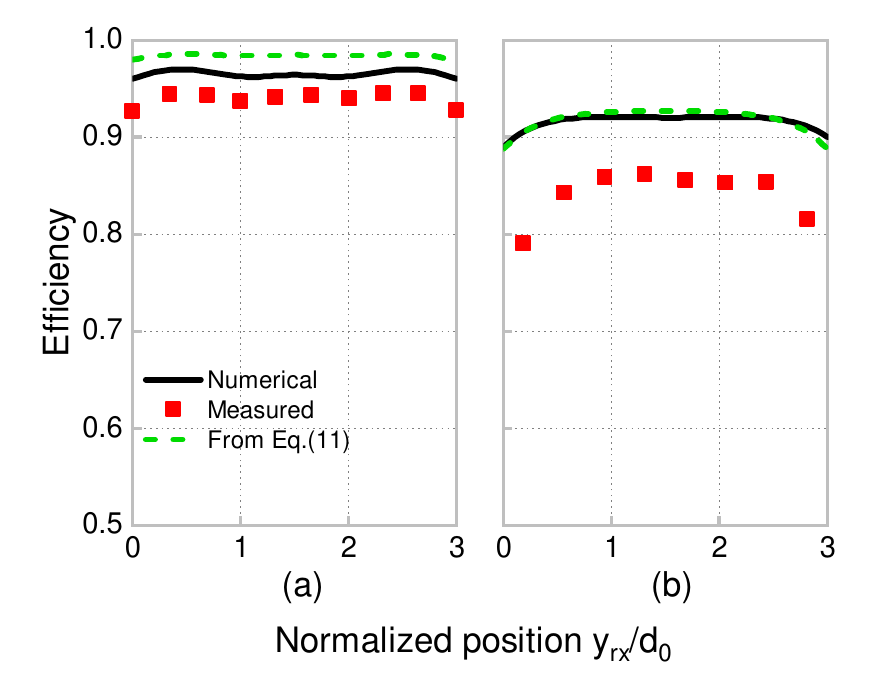}\\
 \caption{Efficiency variation against the normalized position of the receiver when the $z$-distance between repeater array and the receiver is (a)~20 mm, and (b)~50 mm.}\label{measpte}
 \end{center}
\end{figure}

The measured system efficiency variations versus the Rx position $y_{\rm rx}/d_0$ are compared with the numerical calculations in Fig.~\ref{measpte} for $z_{\rm rx}=20$~mm and $z_{\rm rx}=50$~mm. We can see that the numerical calculations using full-form equations (i.e., using \eqref{pte}) is in very good agreement with the simplified expression for the efficiency in \eqref{ptenew2}, which verifies the simplification in \eqref{ptenew2}. The maximum measured efficiency reaches $94.5\%$ for $z_{\rm rx}=20$~mm and $86.2\%$ for $z_{\rm rx}=50$~mm, and its variation is less than $2\%$. The experimental efficiencies are in good agreement with the numerical calculations. The disparity between the measured and numerical results is attributed to additional losses introduced in the experimental setup, such as losses in capacitors and connectors, and a slight mismatch in self-resonance frequencies of the individual coils, as shown in Table~\ref{Table.coil_measure}. Throughout the complete range of the receiver, the efficiency remains constant and at high level without having any active control in the transmitter or receiver circuits.

\begin{figure}
 \begin{center}
 \includegraphics[width=0.6\columnwidth]{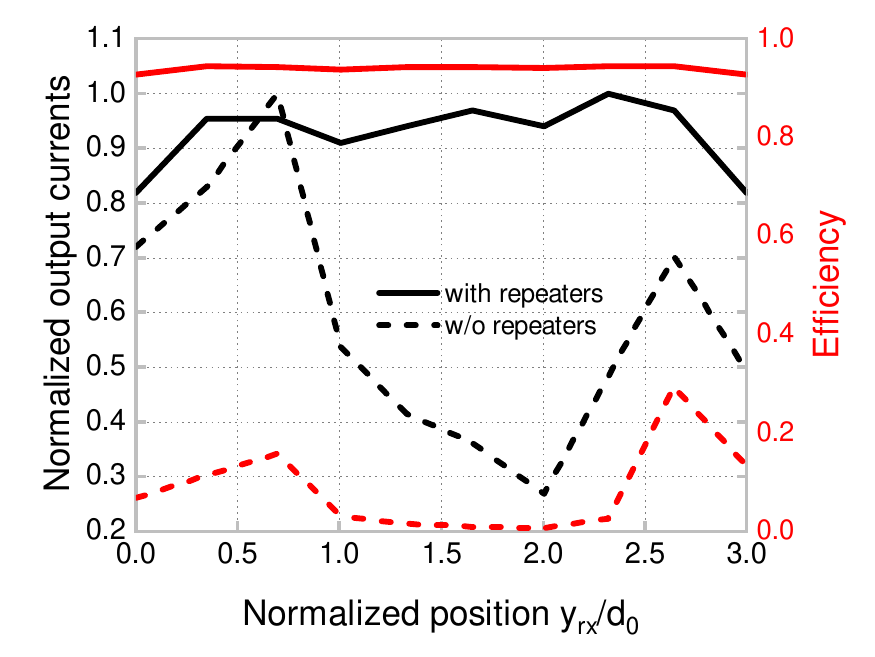}\\
 \caption{The comparison of the output currents and the efficiency of the proposed WPT system with conventional multi-Tx WPT system when the $z$-distance between repeater array to the Rx (for the proposed WPT)  or Tx array to the Rx (for the conventional WPT) is 20 mm.}\label{iout}
 \end{center}
\end{figure}

In order to evaluate the benefits of the proposed method, the characteristics of the proposed approach are compared with a conventional muli-Tx WPT system without the repeater array. The output current and the efficiency variations against the Rx position for the proposed WPT system and the conventional WPT system are compared in Fig.~\ref{iout}. The conventional WPT system  shows great output current variations as compared to the proposed WPT scheme. Most importantly, the efficiency of the conventional approach is very low because of the unoptimized Tx currents. 

\section{Conclusion}
We have presented a WPT system with multiple transmitters and repeaters which automatically tracks the receiver position, always providing an optimized channel for power delivery to freely positioned receivers. This property is achieved due to formation of independent power channels, linking the transmitters and the receiver. These channels gradually switch on and off depending on the receiver position without any control or tuning. The requirements of realizing this feature are 1.~All the cross coupling between channels are negligibly small, and 2.~For a given channel, coupling between the repeater-Rx pair is much higher than that of the Tx-Rx pair. The efficiency, the input and output currents are measured for a system with four Tx coils and four repeater coils. The measured results agree well with the theoretical and the numerical values. The efficiency of the system reaches $94.5\%$ and $86.2\%$ for two different distances between the repeater array and the receiver, $20$~mm and $50$~mm, respectively. Stable and high efficiency is observed for changing receiver location, with the degree of variation less than  $2\%$. Since its simple structure and low cost, the proposed WPT system is a promising solution in many wireless charging applications.


\ifCLASSOPTIONcaptionsoff
 \newpage
\fi

\bibliographystyle{IEEEtran}
\bibliography{IEEEabrv,Bibliography}

\begin{thebibliography}{10}
\providecommand{\url}[1]{#1}
\csname url@rmstyle\endcsname
\providecommand{\newblock}{\relax}
\providecommand{\bibinfo}[2]{#2}
\providecommand\BIBentrySTDinterwordspacing{\spaceskip=0pt\relax}
\providecommand\BIBentryALTinterwordstretchfactor{4}
\providecommand\BIBentryALTinterwordspacing{\spaceskip=\fontdimen2\font plus
\BIBentryALTinterwordstretchfactor\fontdimen3\font minus
  \fontdimen4\font\relax}
\providecommand\BIBforeignlanguage[2]{{%
\expandafter\ifx\csname l@#1\endcsname\relax
\typeout{** WARNING: IEEEtran.bst: No hyphenation pattern has been}%
\typeout{** loaded for the language `#1'. Using the pattern for}%
\typeout{** the default language instead.}%
\else
\language=\csname l@#1\endcsname
\fi
#2}}
\renewcommand\BIBentryALTinterwordstretchfactor{4}

\bibitem{Ahn2013}
D.~Ahn and S.~Hong, ``A study on magnetic field repeater in wireless power
  transfer,'' \emph{{IEEE} Transactions on Industrial Electronics}, vol.~60,
  no.~1, pp. 360--371, Jan 2013.

\bibitem{Hua2019}
R.~Hua and A.~P. Hu, ``Modeling and analysis of inductive power transfer system
  with passive matrix power repeater,'' \emph{{IEEE} Transactions on Industrial
  Electronics}, vol.~66, no.~6, pp. 4406--4413, {Jun} 2019.

\bibitem{Liu2008}
X.~Liu and S.~Hui, ``Optimal design of a hybrid winding structure for planar
  contactless battery charging platform,'' \emph{{IEEE} Transactions on Power
  Electronics}, vol.~23, no.~1, pp. 455--463, Jan 2008.

\bibitem{Jow2013}
U.-M. Jow and M.~Ghovanloo, ``Geometrical design of a scalable overlapping
  planar spiral coil array to generate a homogeneous magnetic field,''
  \emph{{IEEE} Transactions on Magnetics}, vol.~49, no.~6, pp. 2933--2945, Jun
  2013.

\bibitem{Zhong2011}
W.~X. Zhong, X.~Liu, and S.~Y.~R. Hui, ``A novel single-layer winding array and
  receiver coil structure for contactless battery charging systems with
  free-positioning and localized charging features,'' \emph{{IEEE} Transactions
  on Industrial Electronics}, vol.~58, no.~9, pp. 4136--4144, Sep 2011.

\bibitem{Mi2016}
C.~C. Mi, G.~Buja, S.~Y. Choi, and C.~T. Rim, ``Modern advances in wireless
  power transfer systems for roadway powered electric vehicles,'' \emph{{IEEE}
  Transactions on Industrial Electronics}, vol.~63, no.~10, pp. 6533--6545, Oct
  2016.

\bibitem{AbdollahMirbozorgi2014}
S.~{Abdollah Mirbozorgi}, H.~Bahrami, M.~Sawan, and B.~Gosselin, ``{A smart
  multicoil inductively coupled array for wireless power transmission},''
  \emph{IEEE Transactions on Industrial Electronics}, vol.~61, no.~11, pp.
  6061--6070, 2014.

\bibitem{Mirbozorgi2016}
S.~A. Mirbozorgi, H.~Bahrami, M.~Sawan, and B.~Gosselin, ``A smart cage with
  uniform wireless power distribution in 3d for enabling long-term experiments
  with freely moving animals,'' \emph{{IEEE} Transactions on Biomedical
  Circuits and Systems}, vol.~10, no.~2, pp. 424--434, Apr 2016.

\bibitem{Jolani2015}
F.~Jolani, Y.~qiang Yu, and Z.~Chen, ``A planar magnetically-coupled resonant
  wireless power transfer using array of resonators for efficiency
  enhancement,'' in \emph{2015 {IEEE} {MTT}-S International Microwave
  Symposium}.\hskip 1em plus 0.5em minus 0.4em\relax {IEEE}, May 2015.

\bibitem{Kim2012}
J.~W. Kim, H.-C. Son, D.-H. Kim, J.-R. Yang, K.-H. Kim, K.-M. Lee, and Y.-J.
  Park, ``Wireless power transfer for free positioning using compact planar
  multiple self-resonators,'' in \emph{2012 {IEEE} {MTT}-S International
  Microwave Workshop Series on Innovative Wireless Power Transmission:
  Technologies, Systems, and Applications}.\hskip 1em plus 0.5em minus
  0.4em\relax {IEEE}, May 2012.

\bibitem{jaya_optCurrentDistribution}
P.~K.~S. Jayathurathnage, A.~Alphones, D.~M. Vilathgamuwa, and A.~Ong,
  ``Optimum transmitter current distribution for dynamic wireless power
  transfer with segmented array,'' \emph{IEEE Transactions on Microwave Theory
  and Techniques}, vol.~66, no.~1, pp. 346--356, 2018.

\bibitem{Huh2018}
S.~Huh and D.~Ahn, ``Two-transmitter wireless power transfer with optimal
  activation and current selection of transmitters,'' \emph{{IEEE} Transactions
  on Power Electronics}, vol.~33, no.~6, pp. 4957--4967, Jun 2018.

\bibitem{Hui2013}
S.~Y. Hui, ``Planar wireless charging technology for portable electronic
  products and qi,'' \emph{Proceedings of the {IEEE}}, vol. 101, no.~6, pp.
  1290--1301, Jun 2013.

\bibitem{Li2019}
Y.~Li, J.~Hu, L.~Tianren, X.~Li, F.~Chen, Z.~He, and R.~Mai, ``A new coil
  structure and its optimization design with constant output voltage and
  constant output current for electric vehicle dynamic wireless charging,''
  \emph{{IEEE} Transactions on Industrial Informatics}, pp. 1--1, 2019.

\bibitem{pacini2017load}
A.~Pacini, A.~Costanzo, S.~Aldhaher, and P.~D. Mitcheson, ``Load-and
  position-independent moving mhz wpt system based on gan-distributed current
  sources,'' \emph{IEEE Transactions on Microwave Theory and Techniques},
  vol.~65, no.~12, pp. 5367--5376, 2017.

\bibitem{Kim2019}
W.~Kim and D.~Ahn, ``Efficient deactivation of unused {LCC} inverter for
  multiple transmitter wireless power transfer,'' \emph{{IET} Power
  Electronics}, vol.~12, no.~1, pp. 72--82, Jan 2019.

\bibitem{Maslovski2004}
S.~Maslovski, S.~Tretyakov, and P.~Alitalo, ``Near-field enhancement and
  imaging in double planar polariton-resonant structures,'' \emph{Journal of
  Applied Physics}, vol.~96, no.~3, pp. 1293--1300, Aug 2004.

\bibitem{MLcalculations}
Y.~P. Su, X.~Liu, and S.~Y.~R. Hui, ``Mutual inductance calculation of movable
  planar coils on parallel surfaces,'' \emph{IEEE Trans. Power Elect.},
  vol.~24, no.~4, pp. 1115--1123, April 2009.

\bibitem{ADSsoftware}
\BIBentryALTinterwordspacing
{Keysight}, ``{Advanced Design System (ADS)}.'' [Online]. Available:
  \url{https://www.keysight.com/en/pc-1297113/advanced-design-system-ads}
\BIBentrySTDinterwordspacing

\bibitem{kim2014ADS}
J.~Kim, D.-H. Kim, and Y.-J. Park, ``Analysis of capacitive impedance matching
  networks for simultaneous wireless power transfer to multiple devices,''
  \emph{IEEE Transactions on Industrial Electronics}, vol.~62, no.~5, pp.
  2807--2813, 2014.

\end{thebibliography}

\vfill

\end{document}